\documentclass[twocolumn,prl,showpacs,preprintnumbers]{revtex4-1}
\usepackage[applemac]{inputenc}
\usepackage[T1]{fontenc} 
\usepackage{amsmath} 
\usepackage{amsfonts} 
\usepackage{amssymb}
\usepackage{graphicx} 
\usepackage[usenames]{color} 

\usepackage{bbm}
\def\beq{\begin{equation}}
\def\eeq{\end{equation}}
\def\bea{\begin{eqnarray}}
\def\eea{\end{eqnarray}}
\def\ba{\begin{array}}
\def\ea{\end{array}}

\def\l.{\left.}
\def\r.{\right.}

\def\part{\partial}

\begin{document}

\preprint{}
\preprint{}
\title{ON THE INTERACTION OF ELECTROMAGNETIC WAVES WITH CONDUCTORS.}
\author{B.V. Paranjape$^1$}
\email{bhala.jape@gmail.com}
\author{M. B. Paranjape$^{2}$} 
\email{paranj@lps.umontreal.ca}
\affiliation{$^1$Professor Emeritus,
Department of Physics, University of Alberta}
\affiliation{$^2$Groupe de physique des particules, D\'epartement de physique,
Universit\'e de Montr\'eal,
C.P. 6128, succ. centre-ville, Montr\'eal, 
Qu\'ebec, Canada, H3C 3J7 }

\begin{abstract} 
We study the interaction of electromagnetic waves with electrons. Our results can be applied to radio waves in the ionosphere or to lasers impinging on metals causing melting.  We generalize the classical analysis of Zener to the case which includes the interactions of the electrons with lattice vibrations or the positive ions.  { We use the induced polarization to give a globally coherent and unifying analysis of the two cases, where collisions are important and where they are negligible.}
\end{abstract}


\maketitle

\section{Introduction}
Born and Wolf \cite{BornWolf} in their famous book ``Optics'' have devoted a chapter on the ``Optics of metals''.  They have discussed the subject exhaustively, explaining the Drude \cite{Drude} theory of conductivity, taking into account the free electrons, which are free to move inside the metal, except for occasional collisions with lattice vibrations.  The electrons have a finite conductivity as a consequence of these collisions with the lattice vibrations.  Born and Wolf have calculated reflection and transmission coefficients of electromagnetic waves.  

A very simple model of the electrons in metals, which assumes no collisions, was given by Zener \cite{Zener}.  Zener's model is applicable  where the electron density is very low and there are almost no collisions with the positive ions.  Zener's theory is actually well suited for applications to the response of electromagnetic waves in the ionosphere.  He has shown that electromagnetic waves of frequency less than the plasma frequency of the ionosphere are reflected completely.   Radio waves  can be transmitted over long distances on the earth, by reflecting them off the ionosphere.  Zener's theory also explains the qualitative behaviour of electromagnetic waves incident on thin plates of alkaline metals.  It was Kronig \cite{Kronig} who pointed out that Zener's theory will only work well in metals if the frequency of the electromagnetic waves is higher than the frequency of collisions of the electrons with the lattice.

Although Zener's theory is restrictive to certain conditions, it provides a good model for understanding the process, and is useful for students at the undergraduate level.  In the present paper, we extend Zener's theory to include collisions with the lattice vibrations.  Our results are applicable to the case when the frequency of the electromagnetic wave is much lower than the frequency of the collisions of the electrons with the lattice vibrations.  The merit of Zener's approach is in its simplicity; and our approach is just as simple and gives a clear picture of the process.  Thus teachers can use our theory to teach at the undergraduate level.  

In another experimental situation, an intense laser beam can produce local melting in metal plates \cite{Mazur}.  We can conclude from our discussions that the frequency of laser should be lower that the collision frequency, in that way leading to the transfer of energy to the lattice from the laser through collisions of electrons with the lattice vibrations, and induce the melting.  

Early experiments on reflection and transmission of electro-magnetic radiation from thin plates of alkali metals were done by Woods \cite{Woods}.  His observations were explained by Zener, who assumed that the electrons in alkali metals are completely free,  and they respond to the electric field of the incident radiation.  It was observed by Kronig that although the electrons in these metals are free to move inside, they are in fact scattered by lattice vibrations. They are not completely free.  If $\tau$ is the mean time between successive collisions, then the electrons are really free, only for a time $t\ll\tau$; and consequently,  Zener's theory is only good as long as the period of oscillation of the radiation $1/\omega$ is  smaller than $\tau$. That means  $\omega \tau\gg1$. If $\omega \tau\ll1$, then Zener's theory cannot be applied. In the present paper we extend Zener's analysis exactly to the case  $\omega \tau\ll1$.

{ Our analysis presents this interaction through the use of the polarization that the electromagnetic wave induces in the conductor.  We can then present the two cases where collisions are important and where they are negligible in a single, logically coherent and unifying fashion 

\section{Polarization}
When an electric field is applied to a conducting medium, the mobile charges immediately react to the field and rearrange in an effort to cancel it.  In the cases that we shall consider, it is the electrons that move, while the much heavier positive ions are essentially fixed in position.  They may actually be  stuck in place, as in a solid conductor or, simply being much heavier than the electrons, their motion occurs an a much longer time scale and hence is negligible.  

On the other hand even the mobile charges are assumed to be at thermal equilibrium.  Such an equilibrium is only established through the existence of collisions.  However, there is a time scale associated with the collisions and a length scale which depend on the temperature and the density.  Taking a Maxwell-Boltzmann distribution \cite{kd} for the mobile charge carriers, which is a very good approximation for atmospheric plasmas and for metals that are not too cold (so that the electrons are not degenerate), the relations are
\beq
v\sim \left(k_B T/m_e\right)^{(1/2)}\quad l\sim 1/(n\sigma)\quad \tau\sim l/v 
\eeq
where $v$ is the root mean square velocity, $l$ is the mean free path, and $\tau$ is the collision time, which are functions of $k_B$  the Boltzmann constant,  $T$ the temperature, $m_e$ the electron mass, $n$  the  density of the electrons, and $\sigma$  the total scattering cross-section.  But for the Coulomb interaction without screening, the total cross-section diverges, hence it is necessary to take into account the screening of the electric field.  In a conductor, the electric field is screened by rearrangement of the charges.  The Debye length \cite{Debye}, $l_D$,  is the distance over which charge is screened within a conductor.  The derivation of the Debye length is straightforward and yields \cite{ll}
\beq
l_D\sim \left(k_B T/ne^2\right)^{(1/2)}.
\eeq
Calculation of the scattering cross-section crucially requires taking into account the screening length, otherwise the results are divergent.  

Now for an electromagnetic wave, the electric field is oscillating back and forth with  frequency $\omega$, and depending on how high the frequency is, it is important to either neglect collisions or to take collisions into account.  Thus there are two natural regimes, $\omega\tau\gg 1$ and $\omega\tau\ll 1$.  In the former, collisions are unimportant and we treat the electrons in the section below called Free Electrons.  In the latter, the collision are important, and our corresponding section following is called Scattered Electrons.  

In both regimes, the electrons are able to move relative to the heavy or stationary positive ions.  This relative movement induces a local, microscopic dipole moment throughout the conductor,  albeit for a finite amount of time that is proportional the inverse of the frequency of the electromagnetic radiation.   This local, microscopic dipole moment is entirely analogous to the way  an electric field induces a polarization in a dielectric (non-conducting) medium, although the clear difference is that a dielectric can support static electric fields and static dipole moments, while a conductor cannot.  The local, microscopic dipole moment fluctuates wildly over distances comparable with the inter-particle separation, however it is the averaged value of the field produced  that is important.  The analysis is again very straightforward, we refer the interested reader to the classic reference by Jackson \cite{Jackson}.  The result is that there is a polarization that is induced throughout the medium which tries to compensate the applied electric field.  This polarization field, $\vec P$, is given by
\beq
\vec P= n\langle\vec p_{mol}\rangle
\eeq
where $\vec p_{mol}$ is the dipole moment of each microscopic molecule and $\langle\cdots\rangle$ indicates averaging over a macroscopically small volume, which nevertheless contains many microscopic molecules.  The induced electric field is simply $4\pi \vec P$ and the displacement is defined as
\beq\vec D = \vec E + 4\pi \vec P.
\eeq
Computing the average polarization is very easy.  The dynamics of the individual electron motion is governed by Newton's law, the forces acting on the electron are the electric field (the magnetic field also exercises a force on the electrons, but this is down by a power of $v/c$) and the frictional forces if there are  collisions. If $x(t)$ is the electron position (considering for simplicity motion along the $x$ direction), the microscopic dipole moment  is $p_{mol}=-ex(t)$ as the corresponding positive ion is assumed to stay immobile at $x=0$, and the consequent polarization per unit volume  is obtained by multiplying by the number density,  $P=-n e x(t)$.  
}

\section{Free electrons} 

As stated above, Zener' s model is well suited for treating electrons in the ionosphere. The density of electrons in the ionosphere is very low, and the electrons hardly ever collide with the ions.  Then, Zener's approach is most appropriate for dealing with the reflection of electro-magnetic waves.  It is well known that radio signals can be sent across oceans or
over long distances across continents because they can be
reflected off the ionosphere. The ionosphere is a collection of
electrons and an equal number of protons forming a shell around
the earth.  The theory of reflections by free-electrons is easy to
understand: If an electromagnetic wave is incident on the
ionosphere, the electrons which are much lighter than the ions,
are accelerated by the electric field of the electromagnetic wave.

Without loss of generality, take the direction of the electric
field to be along the $x$ direction. All quantities will  have a
periodic variation of angular frequency $\omega$. Thus, the electric
field will be
\beq 
E_x=Ee^{i(kz-\omega t)}.\label{ce}
\eeq
 We also  know this periodic variation exists in all quantities we deal
with as the equations are linear. We will choose  not to write the exponential which is there in all quantities of interest.  {Of course the electric field is actually a real quantity, hence what we mean is the real part of  Equation (\ref{ce}). This does not mean that the periodic behaviour is always necessarily $\cos(kz-\omega t)$, we allow $E$ to be in general a complex number, which then permits any real combination of cosines and sines. Indeed, there may appear explicit factors of $i$ in the subsequent equations,  this simply corresponds to a relative phase difference of $\pi/2$, {\it i.e.} cosines become sines and vice versa.}     

In this electric field, electrons will be
accelerated according to Newton's law, the following equation:
\beq
\ddot x ={{-eE}\over{m}}\label{nl}
\eeq
or
\beq
-\omega^2x={{-eE}\over{m}} 
\eeq
The contribution to
the dipole moment by one electron is then
\beq
-ex={{-e^2E}\over{m\omega^2}}.
\eeq 
If $n$ is the density of
electrons per unit volume, the total dipole moment per unit volume is given simply by
\beq
P={{-ne^2E}\over{m\omega^2}}.
\eeq 
The dielectric
constant at frequency $\omega$ is easily  found through the relation
\beq
\epsilon E=D=E+4\pi P.
\eeq
This gives
\beq
\epsilon(\omega)=1+{{-4\pi ne^2}\over{m\omega^2}}
=1-{{\omega_p^2}\over{\omega^2}},
\eeq 
where
$\omega_p^2={{4\pi ne^2}\over{m}}$, is known as the (square of the)
plasma frequency. The electromagnetic wave equation in a
non-magnetic, isotropic medium is 
\beq
\ddot D=c^2\nabla
^2 E.
\eeq
Which yields
\beq
 -\omega^2\epsilon=-c^2k^2
 \eeq
 that is
 \beq
\omega^2(1-{{\omega_p^2}\over{\omega^2}} ) =c^2k^2. \label{dr}
\eeq
This is called the dispersion relation.  At frequencies less than $\omega_p$, $k^2$ will be negative and hence
$k$ will be imaginary. If $k$ is imaginary, the wave cannot
propagate and it will be totally reflected.

The above analysis can be directly applied to the case of free electrons in
metals. The plasma frequency, $\omega_p$, of the conduction
electrons in metals is very high and much higher than the
frequency of visible light, which is reflected by metals. Thus,
metals look shiny.

\section{Scattered electrons}

We are successful in explaining why the metals shine, but we note that
the free electrons in conducting solids are free only for a short
interval of time. They interact with lattice vibrations and are
scattered after a time $\tau$, the average time between successive
collisions. The above analysis can only be applied to conductors
if the frequency is large $\omega \tau\gg1$.   The electrons can be
accelerated only for a time $\tau$. At low frequencies, the period of oscillation of the electric field is much larger than the time between successive collisions $\tau$.  As a consequence of suffering many collisions, the electrons are not accelerated indefinitely, and acquire a drift velocity
\beq
\dot x={{-eE\tau}\over{m}}. \label{ct}
\eeq
Equation (\ref{ct}) is the defining relation for the collision time $\tau$.  
At low frequencies $\omega \tau\ll 1$. We must replace equation
(\ref{nl}) by equation (\ref{ct}) giving us
\beq
-i\omega x={{-eE\tau}\over{m}}, 
\eeq
that is
\beq
x={{e\tau}\over{i\omega m}}E.
\eeq

Since the
polarization, $P$ is given by 
\beq
P=-nex={{-ne^2\tau}\over{i\omega m}}E,
\eeq
we get
\beq
\epsilon
E=\big (1-{{4\pi ne^2\tau}\over{i\omega m }} \big )E
\eeq
which gives
\beq
\epsilon =1-{{4\pi ne^2\tau}\over{i\omega m}}
\eeq
that is
\beq
\epsilon=1-{{\omega _p^2\tau}\over{i\omega}}.
\eeq
 Define  $\alpha ={{\omega
_p^2\tau}\over{\omega }} $, then
\beq
\epsilon =1+i\alpha.
\eeq 
The dispersion
relation in equation (\ref{dr}) becomes 
\beq
\omega ^2(1+i\alpha)=c^2k^2.
\eeq 
If
$k=k_1+ik_2$, then 
\beq
\omega
^2(1+i\alpha)=c^2(k_1^2-k_2^2+2ik_1k_2).
\eeq 
Equating the real and
imaginary parts, we find
\beq{{\omega ^2}\over{c^2}}=k_1^2-k_2^2,\eeq
\beq{{\omega ^2}\over{c^2}}\alpha=2k_1k_2,~~(k_1^2-k_2^2)\alpha=2k_1k_2;\eeq
\beq k_2^2+{{2k_1}\over{\alpha}}k_2 -k_1^2=0.\eeq
Then we find for the imaginary part
\beq
k_2=-{{k_1}\over{\alpha}}\pm {{\sqrt{{{k_1^2}\over{\alpha
^2}}+k_1^2}}}
\eeq
\beq
=-{{k_1}\over{\alpha}}\pm k_1\Big
(1+{{1}\over{\alpha ^2}}\Big )^{{{1}\over{2}}}
\eeq
\beq 
\approx
{{k_1}\over{\alpha}}\pm k_1\Big (1+{{1}\over{2\alpha ^2}}\Big
)
\eeq
where the approximation is valid for $\alpha\gg 1$ which is the same as assuming a low frequency $\omega$.  
\beq
k_2=k_1\Big (1+{{1}\over{\alpha}}+{{1}\over{2\alpha ^2}}\Big ). \label{ip}
\eeq
The
imaginary part is slightly larger than $k_1$. This indicates that
the wave is absorbed in a distance of less than one wave length,
and does not penetrate into the conductor.

We have chosen to take the $+$ sign of the radical in equation (\ref{ip})
because the negative sign will give us a negative value of $k_2$
which would imply the unphysical result that the wave is amplified. There is no reason
for the amplification. In our analysis, we have introduced collisions of electron which  is a dissipative mechanism. We expect the wave to lose energy
and not gain it.

\section{Discussion and Conclusion}

In conclusion, we observe introduction of scattering and interactions of electrons brusquely impacts the behaviour of electromagnetic wave in conducting media.  The analysis of Zener for the free electron model cannot be applied, specifically in the case $\omega\tau \ll1$.  However a modification of that analysis can be applied to this region taking into account the collisions and the drift velocity that they produce.  The analysis presented here is similar in spirit to Drude's original analysis \cite{Drude}, however there one passes through the conductivity that is induced in the electron current, rather than the more direct route presented here of the polarization that is induced and its effect on the Maxwell equations.  

Thus we see, in a very simple and pedagogical manner, that  at radio frequencies an
electromagnetic wave will travel a great distance along the surface of the earth due to multiple reflections off the ionosphere.  Here the free electron model is sufficient to explain this behaviour.  On the other hand, the same electromagnetic wave will be damped in a real conductor, in which electrons interact with the lattice.  Here the scattering of the electrons with the lattice vibrations cannot be neglected.  Thus, for example,  a superconductor, where electron scattering from the lattice is completely absent,  will reflect radio waves, but if it becomes normal it will suddenly absorb the waves.  {Although it is well understood that superconductivity owes its existence to electron lattice interactions, nevertheless, the electrons in the superconducting state absorb energy, due to interactions with the phonons,  but at an exponentially suppressed rate compared to normal conduction electrons \cite{uas}.  Hence in the superconductor, for temperatures significantly below the transition temperature, the attenuation of ultrasonic waves vanishes.  In this sense, it is correct to interpret this fact simply as the complete transparency of the superconducting electrons to sound vibrations impressed upon the thermal phonon distribution.  

Our presentation makes use of the induced polarization in the conductor to analyze the interaction in the two regimes in a globally coherent, unifying and logical fashion.  We find that this way of presenting the analysis bypasses the necessity of introducing the conductivity, which is only defined without subtlety  in one of the regimes.  In such analyses, the conductivity is computed, and then fed back in to Maxwell's equations to obtain  the dispersion relation \cite{am}.  

The conductivity $\sigma$ is defined by the relation
\beq
\vec J =\sigma \vec E.
\eeq
This relation is the microscopic basis of electrical resistivity.  For a linear, homogeneous conductor of length $L$ and sectional area $A$, we get (taking all vectors in one given direction for simplicity and so dropping the vectorial signs)
\beq
I=J\times A=\frac{\sigma \times A}{L} E\times L=\frac{1}{R} \times V
\eeq
where $I$ is the current, $V$ is the voltage difference and $R$ is the resistivity.  In a collisionless plasma, the resistance is zero, thus the notion of conductivity is subtle.  In fact it can be defined, but turns out to be purely imaginary.  Thus it represents in fact a phase lag between the applied electric field and the corresponding current.   Thus using the conductivity, for example as in \cite{am}, to approach the limit of a collisionless electron gas to arrive at the plasma frequency is not as direct an approach, as we have done in  our section on Free Electrons.

The polarization however is a globally well defined concept in both regimes whether collisions are negligible or important.  Therefore we find that our presentation is simpler, globally coherent and a logical method for exposing and studying the interaction of electromagnetic waves on conductors.
}

\section{Acknowledgements}

We thank NSERC of Canada for partial financial support { and the Perimeter Institute for Theoretical Physics for hospitality where this paper was revised.}


\end{document}